\documentstyle[preprint,aps]{revtex}
\begin{document}

\draft

\title{ 
Multiband theory of quantum-dot quantum wells: Dark excitons, 
bright excitons, and charge separation in heteronanostructures
}
 \author{W. Jask\'olski$^*$
 and Garnett W.  Bryant}

\address{\it National Institute of Standards and Technology,
Gaithersburg, MD 20899, USA\\
e-mail: garnett.bryant@nist.gov}
\date{Current version dated \today }
\maketitle

 \begin{abstract} 
 \noindent 
Electron, hole, and exciton states of multishell CdS/HgS/CdS quantum-dot 
quantum well nanocrystals 
 are determined by use of a multiband theory that includes valence-band 
mixing, modeled with a 6-band Luttinger-Kohn Hamiltonian, and nonparabolicity 
 of the conduction band.
 The multiband theory correctly describes the recently observed dark-exciton   
ground state and the lowest, optically active, bright-exciton states.
Charge separation in pair states is identified. Previous single-band theories
could not describe these states or account for charge separation. 
\end{abstract}
 

\narrowtext

Quantum-dot quantum well (QDQW) nanocrystals
are composed of an internal semiconductor
core which is coated with several shells of different semiconductors           
\cite{mews,schooss}. 
These structures have been synthesized by wet chemistry and can  
have spherical \cite{mews2} or tetrahedral \cite {mews} shape. 
The method of covering  
CdS or HgS nanocrystals by HgS or CdS shells has been established for
several years \cite{e-jl}. Recently, QDQWs containing three 
layers, each with a thickness controlled to a single 
monolayer \cite{mews2,hass}, have been fabricated. 
Transition energies and optical dynamics in these structures can be  
precisely designed by changing the internal core diameter and thickness
of each shell. With the possibility of achieving very uniform size
distributions of dots in a sample and the ability of forming 2D and 3D arrays 
of chemically synthesized nanocrystals \cite{kagan}, 
QDQWs become intriguing candidates as building blocks in QD arrays  
for novel electronic and optical applications.

Recently Mews {\it et al.} \cite{mews} have fabricated and studied QDQW
nanocrystals formed with 4-5 nm diameter CdS cores, 1 ML HgS 
shells, and $\sim$2 nm wide CdS outer cladding layers. Since CdS 
has a wide band gap and HgS has a narrow
band gap, the radial profiles for conduction band 
(CB) and valence band (VB) edges of a CdS/HgS/CdS QDQW each form an internal quantum well 
in the HgS layer. The electron-hole excitations in these structures are determined
by competition between global confinement in the entire nanocrystal, local
confinement in the internal quantum well, and electron-hole pair interaction. 

The low energy optical excitations in these QDQWs have been measured by absorption,
luminescence, flourescence line narrowing (FLN), and hole burning (HB) \cite{mews}. The lowest
optically active electron-hole pair state is separated from the next optically 
active pair state by about 60 meV. A large, 19 meV Stokes energy shift 
is observed between
the lowest optically active pair state, that is used as the excitation level in the 
flourescence measurements, and the main emission peak, indicating
that the ground state is a {\it dark} exciton.

Electron, hole, and exciton states
of QDQWs have been investigated so far only with the one-band
effective mass approximation, treating a light hole with a
mass similar to the CB mass \cite{schooss,mews2,gwb}.  The energy
of the main absorption peak can be predicted reasonably well by
these calculations. However, in these calculations, the main 
absorption peak arises from transitions to the
lowest pair state, there is no dark-exciton ground state. Also, 
the next optically active pair state is predicted to be 200 meV 
above the lowest optically active state. Since the electron and
hole have similar masses in these models, little separation of
the electron and hole into different layers is predicted.

The presence of multiple, closely spaced excitations with very different oscillator
strengths suggests that a more detailed description of the band states, including
both heavy and light holes, is needed for these QDQWs. It has been proved for 
other semiconductor quantum
dots \cite{grigorian,ekimov,n-b,wj} that valence-band mixing must
be included to correctly describe hole levels, transition energies, and excitation
spectra. For structures containing layers of narrow-gap semiconductors,
such as HgS, CB nonparabolicity should also be included.
To explain the recently observed spectra of CdS/HgS/CdS quantum
dots \cite{mews}, to determine when charge separation occurs, and to study 
how energy levels and excitation spectra
depend on CdS and HgS shell thicknesses, 
we have performed  multiband calculations for spherical QDQWs based
on the $\rm \bf k \cdot p$ method and the envelope function approximation (EFA). 

We use the 6-band Luttinger-Kohn Hamiltonian in
the spherical approximation \cite{grigorian} to describe hole states. 
Only the angular momentum operator $F=J+L$, where $J$ is the Bloch band-edge
angular momentum (3/2 for heavy and light holes and 1/2 for the split-off band)
and $L$ is the envelope angular momentum in a spherical dot, commutes
with the hole Hamiltonian. The hole states are eigenfunctions of $F$ and $F_z$ 
\begin{equation}
|FF_z;nL^h\rangle =\sum_{J,L \geq L^h}\sum_{J_{z},L_{z}}\langle JJ_zLL_z;FF_z\rangle |JJ_z\rangle 
|nLL_z\rangle 
\end{equation}
where the 
$|JJ_z\rangle $ are the appropriate Bloch band-edge states, 
$\langle {\rm \bf r}|nLL_z\rangle =f_{nL}(r)Y_{LL_z}({\rm \bf \hat r})$, the 
$f_{nL}(r)$ are radial envelope functions and the $Y_{LL_z}({\rm \bf \hat r})$ are
spherical harmonics. Following Ref. \cite{ekimov,wj} the hole states
are described by three quantum numbers: $nL^h_F$, where $n$ is the
main quantum number, and $L^h$ is the lowest $L$ that appears in
Eq.(1) for a given $F$. The three different radial components $f_{nL}(r)$  
that appear for a given $F$ are solutions of a set of second-order
coupled differential equations for the radial part of
the 6-band Luttinger-Kohn Hamiltonian. For each semiconductor shell 
this Hamiltonian depends on 3 empirical parameters:
two Luttinger parameters $\gamma$ and $\gamma_1$
and the split-off gap $\Delta$.

The electron states are products of the
Bloch CB-edge state $|S\sigma \rangle $ for an $S$ atomic state with
spin $\sigma$ and the envelope functions $|nL^e L_z^e\rangle $.
The one-band effective-mass radial equation is solved to determine
$f_{nL^e}(r)$. CB nonparabolicity is included perturbatively by use 
of an energy-dependent mass correction defined by two empirical parameters: 
the energy gap $E_g$ and $E_p=2V^2$, where $V=\langle S|p_z|Z\rangle $ is  
the Kane matrix element \cite{ekimov}. 
The electron and hole equations are solved numerically. 

We use the following material parameters:
CdS $E_g=2.5$ eV, $\gamma_1=0.814$, $\gamma=0.307$, $\Delta=0.08$ eV, $E_p=19.6$ eV;
HgS  $E_g=0.2$ eV, $\gamma_1=12.2$, $\gamma=4.5$, $\Delta=0.08$ eV, $E_p=21.0$ eV.
The heavy and light hole masses resulting from the  
$\gamma$, $\gamma_1$ are: CdS $m_{hh}=5.0$, $m_{lh}=0.7$; HgS 
$m_{hh}=0.31$, $m_{lh}=0.047$. 
These parameters are the same as or close to the corresponding values found in
the literature \cite{hw,biel,l-b,bp,comment1}. The HgS band gap is taken to be
positive. This is consistent with recent measurements of the HgSe band gap \cite{gawlik}.
Based on the photoelectric thresholds for CdS and HgS \cite{neth},  
the CB and VB offsets are taken to be $1.45$ eV and $0.85$ eV, respectively, which are close
to the values used in previous calculations \cite{schooss,mews2,gwb}. 
The barriers for tunneling into water, which is the medium 
surrounding the QDQW, are $4$ eV for both electrons and holes 
(photoelectric threshold in H$_2$O 
is $\approx 8$ eV) when measured from the middle of the HgS gap. 
The H$_2$O masses are $m_{hh}=m_{lh}=m_{e}=1.0$ ($\gamma_1=1.0$ 
and $\gamma=0.0$) \cite {gwb}. The choice of H$_2$O masses is
not critical since the high H$_2$O barriers prevent any 
significant leakage out of the QDQW.
Contribution from higher electronic bands is taken into account
by use of the parameter $f=-1.0$ in the electron effective mass equation \cite{ekimov}.
As a result the electron mass is 0.15 near the CB-edge in CdS and 0.04 for the energy   
range of interest in HgS, close to values found in the literature \cite{gwb,biel,l-b}.

To test that the EFA can be applied to QDQW structures containing
layers as thin as 1 ML, we perform first a series of calculations for 
wide-layer structures, for which the EFA works   
\cite{ekimov,n-b,wj}, and then we vary shell widths to reach the 
limit of thin layers. The sequence is shown in Fig. 1.
We start with a CdS quantum dot with a 2 nm radius, {\it i. e.} a 1 nm 
core and a 1 nm clad (structure {\it a} in Fig. 1).
Next, we add a HgS shell between the CdS core and clad, starting with a
0.3 nm ($\sim$1 ML) shell and extending to a 2 nm shell
(structure {\it b} in Fig. 1). Next the CdS core is reduced
until the limit of a HgS/CdS quantum dot with no CdS core 
(structure {\it c}) is reached.  Finally, the 1 nm wide CdS clad is eliminated to 
end with a 4 nm diameter HgS QD (structure {\it d}). 

Electron and hole energies are shown in Fig. 2 for this sequence
of structures. Transition energies are calculated by taking the
electron-hole pair energy differences and subtracting the pair binding energy, 
which is determined perturbatively \cite{brus} with an 
average effective dielectric constant. The transition energies are presented in Fig. 3a. 
Oscillator strengths of the lowest transitions are
shown in Fig. 3b. The oscillator strengths are calculated by averaging over all
linear polarizations ({\it i}) of the dipole transition operator
\begin{equation}
\sum_{L^{e}_{z}\sigma} \sum_{i}|\sum_{F_{z} 
J_{z}}\langle JJ_zLL_z;FF_z\rangle 
\langle n^eL^eL^e_z|n^hLL_z\rangle \langle S\sigma |\hat{p}_i|JJ_z\rangle |^2
\end{equation}
where $\sum_{L^{e}_{z}}$ averages over final electron states and
$\hat{p}$ is the momentum operator. Most importantly, electron 
and hole levels evolve smoothly as layer thicknesses
are varied. Thus the EFA should be quantitatively accurate for structures
with wide layers and should be qualitatively accurate and quantitatively 
reasonable for structures with thin layers.

As the HgS shell width increases, successive electron states become trapped in the
HgS shell when their energies fall below the CdS CB-edge and their charge
densities become localized in the HgS shell. Due to global confinement,
electron energies increase as the CdS core or clad decreases.

In the one band approximation \cite{schooss,mews2,gwb}, hole and electron 
states behave the same way when the HgS thickness is varied, with the 
corresponding hole and electron states trapping in the HgS for nearly 
the same thickness. In the multiband approximation,
hole states behave differently from electron states. 
A group of hole levels ($1P_{3/2}$, $1P_{1/2}$, $1S_{3/2}$) 
{\it easily} fall below the CdS VB-edge, even for a HgS shell as thin as
1 ML ($\sim 0.3$ nm). 
The corresponding charge densities are strongly localized
inside the HgS (see Fig. 4). These hole states are more easily trapped than the corresponding
electron states. The $n=2$ hole states of these symmetries
trap in the HgS layer at larger widths ($\sim 0.4-0.5$ nm). 
There is also a group of states ($nS_{1/2}$) with energies above the CdS
VB-edge even for a 2 nm wide HgS shell. Their charge density maxima are located
in the CdS cladding layer (see Fig. 4).

The $1S_{1/2}$ hole state does not trap in the HgS layer because the 
CdS and HgS hole effective masses are so different.
The $1S_{1/2}$ state is made from light hole and split-off bands
only. The HgS light hole and split-off band masses  
are about 15 times lighter than the corresponding CdS masses.
The HgS shell acts as a barrier for the $S_{1/2}$ state, rather
than a potential well, because the hole
has such a light mass and high kinetic energy in that shell.
Moreover, the dominant contribution from the $J=3/2$ band 
to the lowest $S_{1/2}$ state is made by the
$L=2$ component \cite{grigorian,n-b}. Thus the $S_{1/2}$ charge density maximum
is in the CdS clad. In contrast, $S_{3/2}$, $P_{3/2}$ and $P_{1/2}$ 
states are a 
mixture of heavy hole, light hole and split-off bands. 
The HgS heavy hole mass 
and the CdS light hole mass are similar, so these states can localize 
in the HgS well. 

Localization of the hole $1S_{1/2}$ state in the CdS clad
and electron $1S$ state in the HgS shell (see Fig. 4) 
explains why the oscillator strength (Fig. 3b) of the
$1S_{1/2}-1S$ transition is about two orders of magnitude smaller than of
$1P_{3/2}-1P$ or $1P_{3/2}-1P$ transitions. Including the effects
of pair interaction beyond the perturbation energy shift included in our
calculations would not dramatically reduce this charge separation because
quantum confinement effects dominate pair binding in these small
structures \cite {gwb}.
The binding energy of an exciton in this pair state should be smaller than
in other pair states, since the
electron and hole are strongly localized in different layers. 
The oscillator strength of the $1S_{3/2}-1S$ transition is even
smaller than for the $1S_{1/2}-1S$ transition. 
Only the $L=0$ component  
of $1S_{3/2}$ state yields to a non-vanishing transition dipole.
In these QDQW structures this component of the hole state  
is negligible compared to the two $L=2$ components.
As a consequence, the $1S_{3/2}-1S$ transition is  
optically inactive. 

Finally, we perform specific calculations for the QDQW nanocrystals investigated
recently by Mews {\it et al.} \cite{mews}. We consider a structure with a
CdS core of radius 2.2 nm, 1 ML (0.35 nm) HgS shell and
2 nm wide CdS clad. The calculated energies of the 
first two optically allowed transitions, 
$1P_{3/2}-1P$ and $1P_{1/2}-1P$, 
are 1.890 eV and 1.929 eV, respectively. 
In the hole burning experiment by Mews
{\it et al.} \cite{mews,priv} the first excitation energy peak appears at
1.878 eV, only 12 meV different from the calculated energy of the 
lowest $1P_{3/2}-1P$ electron-hole pair state.
The next experimentally observed  excitation 
is $\sim60$ meV higher (Fig. 2 of 
Ref. \cite{mews}) and differs from the predicted position
of $1P_{1/2}-1P$ state by only 10 meV.
Experimentally, both transitions should be
of  comparable strength \cite{priv}. Our calculated
oscillator strengths are almost the same for these transitions (see Fig. 3b).
The calculated energy of the optically inactive 
$1S_{3/2}-1S$ transition is red-shifted from 
the ground $1P_{3/2}-1P$ transition
by 18 meV in almost perfect agreement with the 
19 meV difference observed 
between absorption and emission peaks in 
FLN (Fig. 2 of Ref. \cite{mews}).
Thus the $1S_{3/2}-1S$ pair state is the 
{\it dark exciton} in this structure. 
The energy shift $\Delta_{HB-FLN}$ between {\it excitation} 
peaks of HB and FLN spectra shown in Fig.2 
of Ref. \cite{mews} most likely occurs because the samples contain 
a distribution of QDQWs. 
The lowest calculated 
transition redshifts approximately by $\Delta_{HB-FLN}$ when
the CdS core radius increases to 2.8 nm.

In conclusion, a multiband theory of electron, hole, and exciton states in
QDQWs has been developed.
Multiband calculations show that for some pair states, the electron and
hole can be trapped in different shells, yielding weak
oscillator strengths for these transitions.
Other transitions can be weak even if both
electron and hole are localized in the same layer. 
The observed energy difference between the lowest optically active
transitions in CdS/HgS/CdS nanocrystals, as well as the appearance of 
{\it dark exciton} can be explained by the multiband theory. 
This could not be done in the one-band approximation.
These results show that the EFA can be applied to interpret
optical spectra of nanostructures 
containing layers as thin as 1 ML. For even more accurate description,
corrections due to any nonspherical shape of the dots, pair exchange and  
correlation should be included as well.\\

\noindent
{\bf Acknowledgments}\\

Support from  The Fulbright Foundation and 
KBN project No. 2 PO3B 156 10 is gratefully acknowledged.

\begin{figure}
\caption{ 
The sequence of CdS/HgS/CdS quantum-dot quantum wells investigated and  
the corresponding schematic layout of CB and VB edges.  
}
\label{Fig1}
\end{figure}

\begin{figure}
\caption{ 
The lowest electron (a) and hole (b) energy 
levels for the sequence of structures shown in Fig. 1.
Left part: HgS shell increases from 0 nm to 2 nm (from left to right);
middle: CdS core decreases from 1 nm to 0 nm; right: CdS clad decreases from 1 nm to 0 nm.
}
\label{Fig2}
\end{figure}

\begin{figure}
\caption{ 
Transition energies (a) and oscillator strengths (b) for the 
QDQW sequence in Fig. 1.
}
\label{Fig3}
\end{figure}

\begin{figure}
\caption{
Charge densities of several electron and hole states for a CdS/HgS/CdS QDQW
with CdS core, radius 2.2 nm, 0.35 nm HgS shell and 2 nm CdS clad. 
Vertical bars mark the HgS shell.
}
\label{Fig4}
\end{figure}



\begin{thebibliography}{99}
\bibitem{comment1} Not all of these parameters are well known, particularly for
HgS. Our choices are best estimates from the available values.
\end{thebibliography}
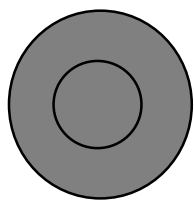
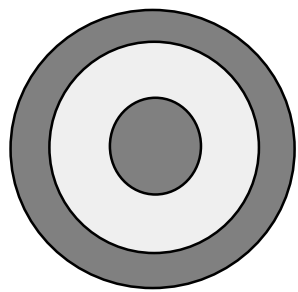
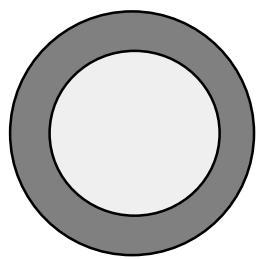
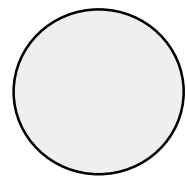
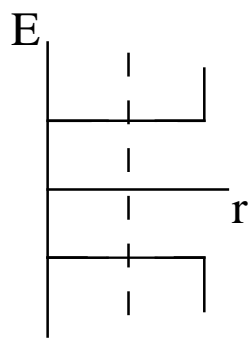
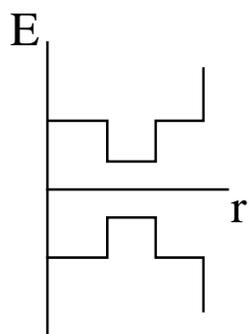
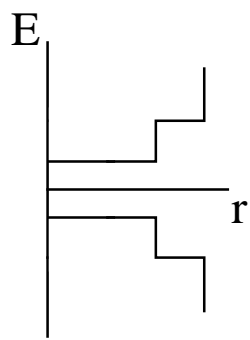
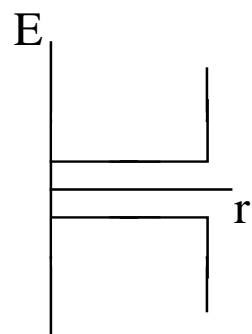
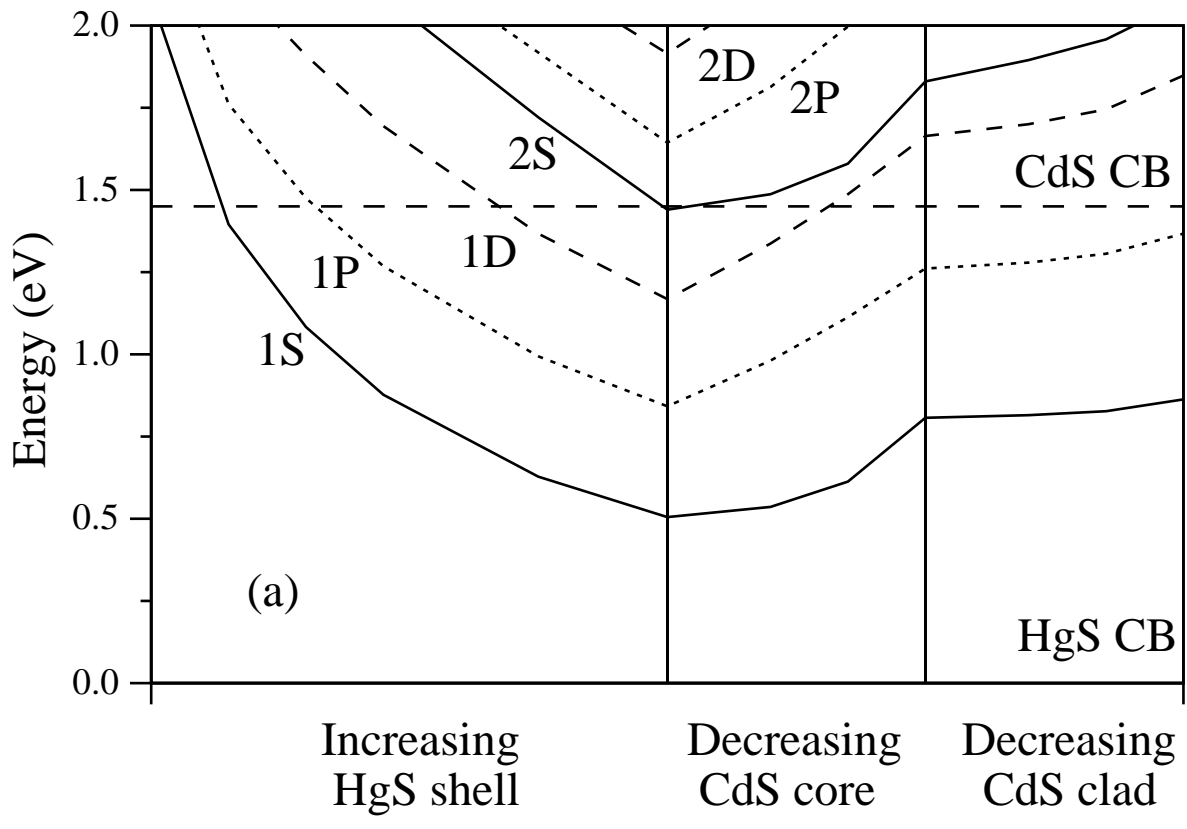
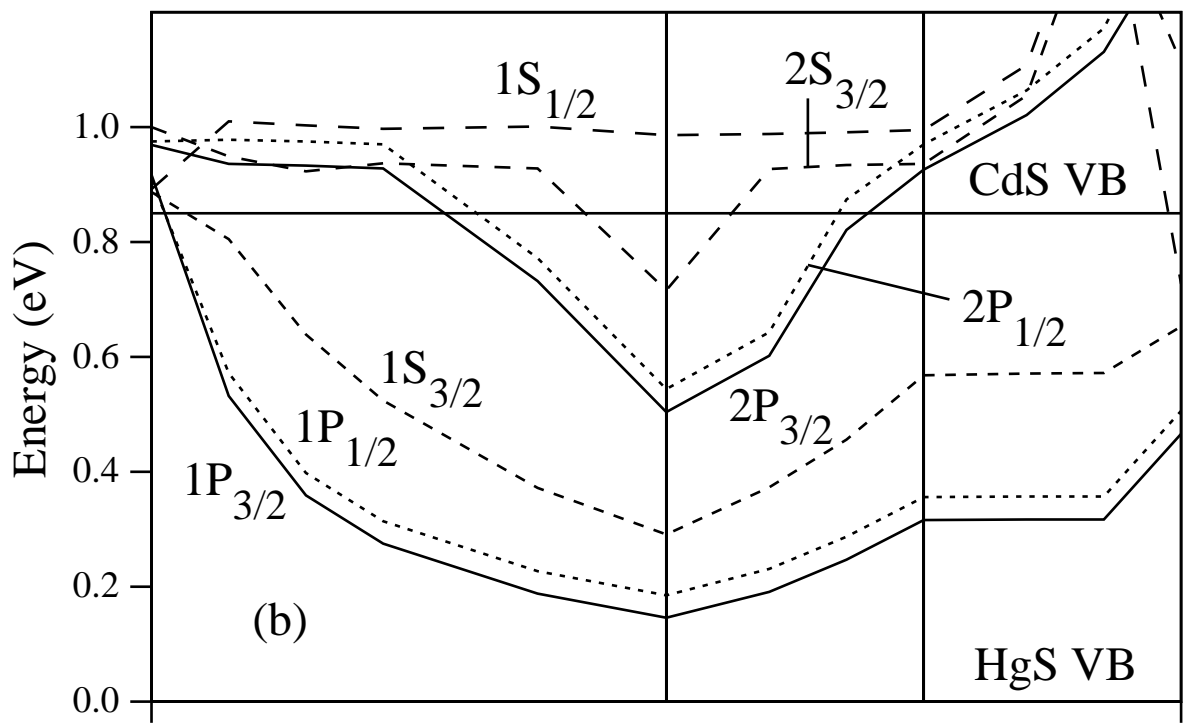
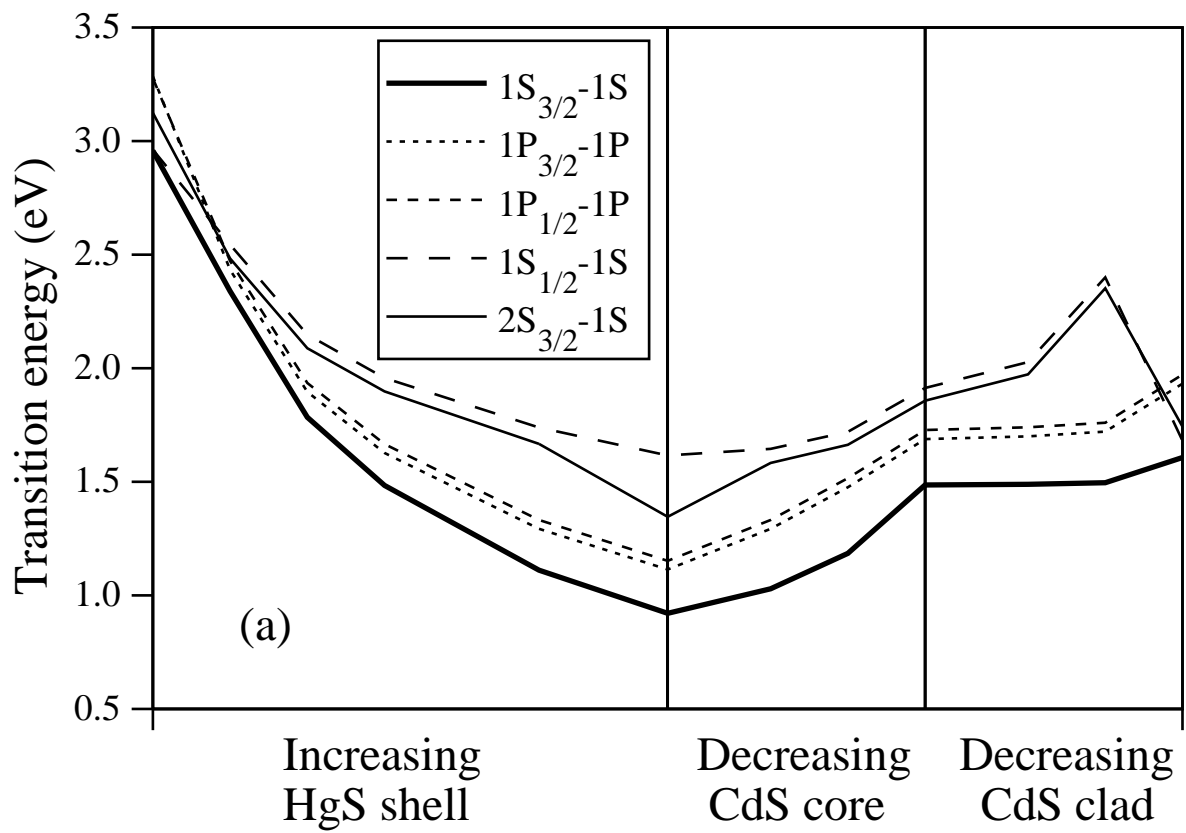
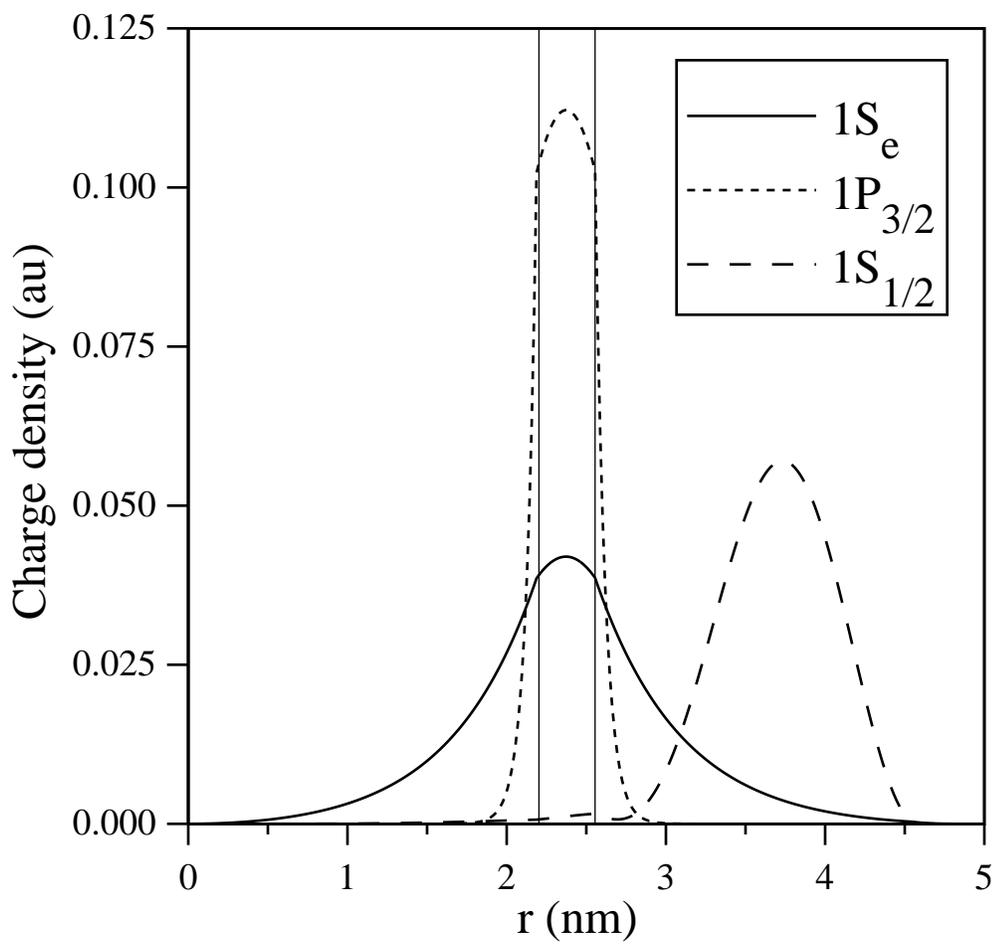
\end{document}